\documentclass[fleqn,11pt]{wlscirep}
\usepackage[utf8]{inputenc}
\usepackage[T1]{fontenc}
\usepackage{bm}
\usepackage{hyperref,xcolor,lineno}
\usepackage{pdflscape}
\usepackage{aps_macros}
\usepackage{enumitem}
\usepackage{ulem}

\graphicspath :
\graphicspath{{figures/}}

\title{High-energy neutrino transients and the future of multi-messenger astronomy}

\author[1]{Claire Gu\'epin}
\author[2,3]{Kumiko Kotera}
\author[4,*]{Foteini Oikonomou}
\affil[1]{Joint Space-Science Institute, University of Maryland, College Park, MD 20742, USA}
\affil[2]{Sorbonne Universit\'{e} et CNRS, UMR 7095, Institut d'Astrophysique de Paris, 98 bis bd Arago, 75014 Paris, France}
\affil[3]{Vrije Universiteit Brussel, Physics Department, Pleinlaan 2, 1050 Brussels, Belgium}
\affil[4]{Institutt for fysikk, NTNU, Trondheim, Norway}

\affil[*]{e-mails: cguepin@umd.edu, kotera@iap.fr, foteini.oikonomou@ntnu.no}

\begin{abstract}
The recent discovery of high-energy astrophysical neutrinos and  first hints of coincident electromagnetic and neutrino emission herald the beginning of the era of multi-messenger astronomy. Due to their high power, transient sources are expected to supply a significant fraction of the observed energetic astroparticles, through enhanced particle acceleration and interactions. Here, we review theoretical expectations of neutrino emission from transient astrophysical sources and the current and upcoming experimental landscape, highlighting the most promising channels for discovery and specifying their detectability.   

\end{abstract}
\begin{document}

\flushbottom
\maketitle

\thispagestyle{empty}

\section{Introduction}\label{sec:introduction}

Transient emissions are short (up to months), violent, and irregular emissions, that could be isolated events or appear as a sudden activity increase of an existing source. Many new classes of luminous transients have been discovered recently, thanks to improved instrumental sensitivity, time resolution, and fast follow-ups across the full electromagnetic spectrum. Many more discoveries are expected with the launch of wide field instruments, multi-instrument platforms and multi-messenger (MM) networks. 
In 2017, a MM window opened on the high-energy (HE) transient sky, with the detection of gravitational waves from the merger of two neutron stars (NS), GW170817, jointly with electromagnetic counterparts\cite{Abbott_2017}. HE ($\gtrsim 10^{15}\,$eV) and ultra-high energy (UHE, $\gtrsim 10^{17}\,$eV) neutrino astronomy should be the next window to open, likely with the MM observation of a flaring neutrino transient source. 

Because they can inject their huge amount of energy over short timescales, powerful transients are the most promising sources capable of supplying enough energy and flux to astroparticles at the observed level. The increase in luminosity can be related to enhanced cosmic-ray acceleration, with subsequent particle interactions and production of secondary emissions, such as HE and UHE neutrinos and gamma-rays. Moreover, anisotropy, source-density, energetics and magnetic-structure arguments strongly challenge steady-source scenarios for UHE cosmic rays with light composition\cite{2019FrASS...6...23B, Abreu13_18, Anchordoqui:2018qom, 2016ApJ...832L..17F, Palladino_2020}. By observing the transient sky at the highest energies, we thus naturally tackle the long-standing enigma of the origin of the highest energy astroparticles.

HE and UHE neutrinos are expected to play a major role in precision time-domain astronomy, as undeflected signatures of hadronic interactions. 
Their deeper cosmological horizon compared to cosmic rays and gamma rays implies that they can be observed from sources located at cosmological distances, possibly in coincidence with gravitational waves. Neutrinos, like secondary photons, can also exhibit time-variability signatures, which can be combined with other wavelengths to probe the structure and evolution of the source and its environment. 

From an experimental point of view, the detection of a transient HE or UHE neutrino event is favored, as time-dependent neutrino searches tend to reduce the background (e.g., due to atmospheric neutrinos and muons in IceCube or ANTARES\cite{2015ApJ...807...46A, ANTARES2015}). Real-time analysis and MM follow-up of alerts can also drastically increase the statistical significance of associations. The detection of several neutrinos from a single source  (``neutrino point sources'') is the ultimate goal of HE neutrino astronomy. What constitutes a detected neutrino point source varies from detector to detector and depends on the neutrino background at the energy and arrival direction of each individual event. In general, an accumulation statistically inconsistent with an accumulation of atmospheric or astrophysical neutrinos by chance is needed in order to assertain the detection of a true astrophysical point source.
From an astrophysical perspective, the detection of neutrino point sources (or small scale clustering referred to as ``multiplets'') depends upon the overall number density of sources\cite{2016PhRvD..94j3006M}. At UHE, currently known powerful steady sources are too rare to enable the detection of multiplets, whereas transient sources could lead to a burst of neutrinos\cite{Palladino_2020}. A guaranteed background for astrophysical neutrino searches at UHE is the flux of neutrinos produced in the interactions of ultra-high energy cosmic rays (UHECRs)~\cite{Berezinsky:1969erk}, often referred to as ``cosmogenic'' neutrinos. These two distinct diffuse astrophysical neutrino populations (cosmogenic versus astrophysical neutrinos produced in source) will be difficult to distinguish without exceptional statistics but will, once discovered, give complementary information about the sources of UHECRs. Point sources of UHE neutrinos, would reveal in-source neutrino production, as the deflections and time-delays suffered by UHECRs make the detection of point sources of cosmogenic neutrinos unlikely.

IceCube has been accumulating diffuse HE neutrino data since 2008\cite{2013PhRvL.111b1103A, 2020PhRvL.124e1103A} and has been sending and receiving alerts via large-scale networks~\cite{2017APh....92...30A}. These searches have led to the detection of a neutrino of energy $\sim 0.1-1\,$PeV, IceCube-170922A, in association with a flaring blazar, TXS 0506+056, observed in multiple wavelengths, including in X-rays and gamma-rays\cite{TXS2018}. The theoretical and experimental interpretations of this association are still under debate, and await confirmation with further observations. Several other possible coincident detections have been discussed\cite{2016GCN..19381...1S, Kadler2016, Stein20}.

Theoretical models of time-variable neutrino emission from luminous transient sources are being developed. The models are based on analytical and numerical approaches, with various levels of complexity in modeling the parent cosmic-ray acceleration and escape/interaction regions in and surrounding the sources\cite{2019ARNPS..69..477M}.
Thanks to theoretical and experimental progress, we are starting to place strong constraints on the structure and energetics of powerful sources that are {\it not} detected in HE neutrinos.

Many neutrino observatories from HE to UHE are in construction or planned, with increased sensitivity, optimized fields of view, and for some of them, excellent angular resolution. The diverse envisioned designs should help tackle different aspects of the challenging detection of neutrino transients. One of the goals of this review is to summarize the theoretical 
ingredients in the numerous sources models as well as the 
characteristics of current and future instruments, and to connect both ends to assess MM detection performances.
This could help refine models, optimize instrumental designs and call for synergies between projects. 

\begin{figure}[!th]
\centering
\includegraphics[width=0.65\linewidth]{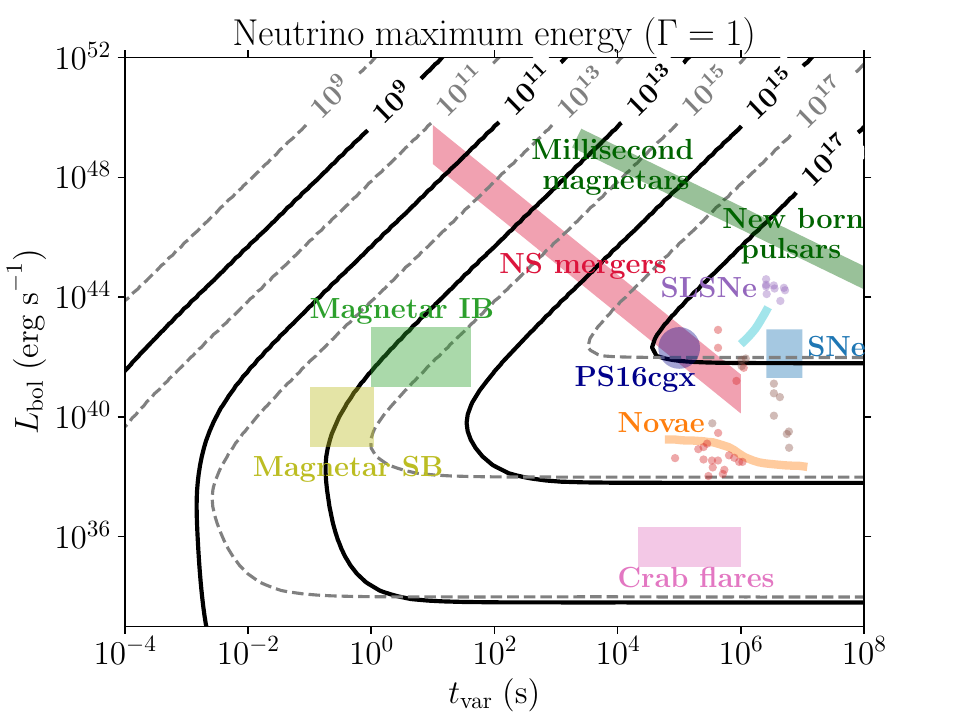}
\includegraphics[width=0.65\linewidth]{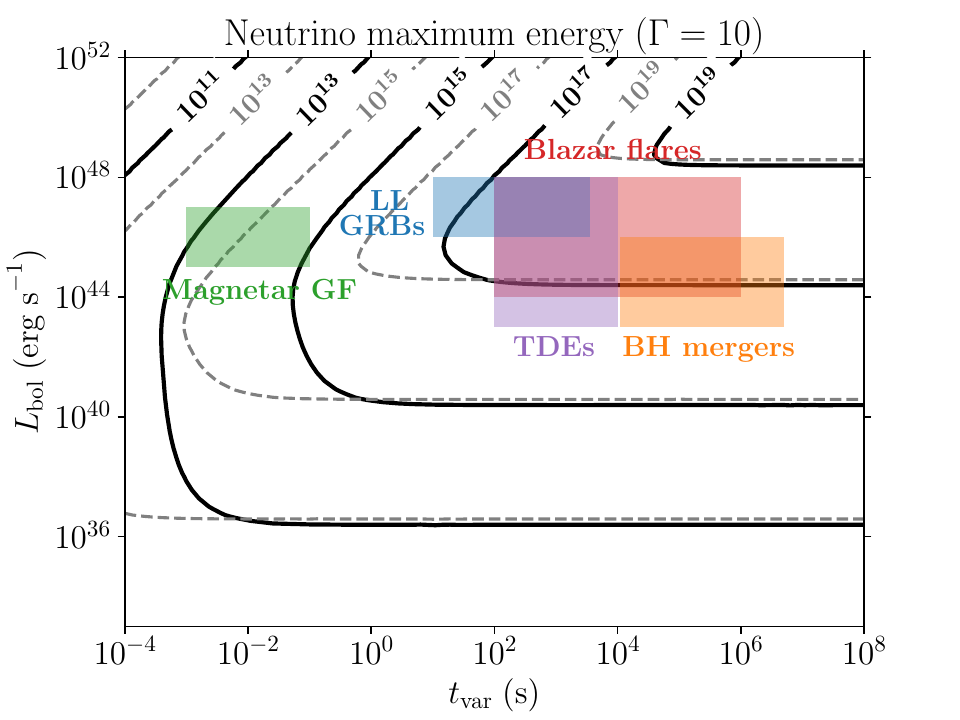}
\caption{Maximum neutrino energy \cite{Guepin17} (in eV) that could be produced by various categories of transient sources, with variability timescale $t_{\rm var}$ and bolometric luminosity $L_{\rm bol}$. We represent non-relativistic sources with outflow Lorentz factor $\Gamma = 1$ {\it (top)} and relativistic sources with $\Gamma = 10$ {\it (bottom)}. 
Neutrinos directly produced through pion (muon) decay are shown with dashed (solid) lines. Overlaid are the properties of different categories of transient sources (colored regions).}
\label{fig:hillas}
\end{figure}

\section{Estimating neutrino fluxes from transient sources: methodologies}\label{sec:modeling}

HE neutrinos are produced through the interaction of accelerated cosmic rays with photon or baryon backgrounds. Various interaction channels are involved, such as the production of charged pions, charged kaons or charm hadrons and their subsequent decay (for instance $\pi^+ \rightarrow \mu^+ +\nu_\mu$ and $\mu^+ \rightarrow e^+ + \bar{\nu}_\mu +\nu_e$), or nuclear decays.
In the following sections, we discuss how regions involved in HE neutrino production are modeled (section~\ref{sec:modeling_regions}), and how the interplay of acceleration and interaction timescales (section~\ref{sec:modeling_timescales}) allows us to estimate neutrino fluxes from transient sources from phenomenological (section~\ref{sec:modeling_pheno}) and numerical (section~\ref{sec:modeling_numerical}) points of view. 

\subsection{Modeling cosmic-ray acceleration and interaction regions}\label{sec:modeling_regions}

In the case of transient sources, interactions leading to the production of transient HE neutrino signals involve backgrounds located inside the sources or in their vicinity. Therefore, most transient neutrino signatures are governed by the interplay between typical timescales that characterize the acceleration, interaction and escape of cosmic rays in these regions. Transient source models usually involve one or more zones inside the source, where different processes can take place. In one zone models, particles are accelerated, propagate and interact in a region that produces all observed emission. When this type of description is inadequate, for instance when one zone models fail at explaining multi-wavelength or MM data\cite{Xue19}, two zone models can be adopted. For instance, acceleration can be treated prior to interactions (first zone) and accelerated  hadrons and leptons can be injected in the interaction region (second zone) where they only propagate and interact. Another example is the use of an external radiation background as a target for interactions. The use of two-zone models should be well motivated by a sufficient amount of available multi-wavelength or MM data, due to the caveat that considering two or more zones allows for additional degrees of freedom.

\subsection{Typical timescales involved in transient HE neutrino production}\label{sec:modeling_timescales}

Acceleration and interactions of cosmic rays operate at microscopic scales that are not accessible to macroscopic source models. The modeling of the typical timescales characterizing these processes allows to evaluate how they compete with each other and thus estimate key quantities such as the maximum energy and flux of neutrinos. These timescales are usually calculated in the comoving frame of the acceleration or interaction region. For various categories of transient sources, the dominant processes are cosmic-ray acceleration, synchrotron and adiabatic losses, and photopion production. Other processes can thus often be neglected at first order. In the following, quantities in the comoving frame are denoted by primed quantities. Other quantities are in the observer frame.

A precise description of acceleration processes, such as shock acceleration \cite{Marcowith16} or magnetic reconnection \cite{Yamada10} for instance, requires the modeling of the astrophysical plasma and the complex interplay between the charged particles and the electromagnetic fields, at scales of the order of the plasma skin depth.
Detailed plasma simulations \cite{Zhang:2021akj} and theoretical estimates \cite{Lemoine19} can be used to derive estimates of the acceleration timescale $t_{\rm acc}$ and the spectrum of accelerated particles, often modeled by a power-law with maximum energy $E_{\rm max}$ and spectral index $\alpha$. Various acceleration processes involve the scattering of charged particles on magnetic inhomogeneities, thus the acceleration timescale can be related to the particle Larmor time $t'_{\rm L} = E' / (c Z e B')$
\begin{equation}
    t'_{\rm acc} = \eta_{\rm acc}^{-1} \, E'(c Z e B')^{-1} \, ,
\end{equation}
where $E'$ and $Ze$ are the energy and charge of the particle, and $B'$ the mean magnetic field. In most cases $\eta_{\rm acc}^{-1} \gg 1$ \cite{Lemoine09} and $\eta_{\rm acc}^{-1} \sim 1$ could be achieved for a maximally efficient energy dissipation mechanism.

The particles are typically confined in the transient source during a dynamical timescale $t_{\rm dyn}$, which characterizes the evolution of the radiation region. This is a central quantity in the modeling of flaring transients and can be related to the observed variability timescale of the emission $t_{\rm var}$ by a condition of causality
\begin{equation}
   t'_{\rm dyn} = (1+z)^{-1} \delta \, t_{\rm var} \, ,
\end{equation}
where $z$ is the redshift of the source and $\delta = [\Gamma(1-\beta \cos\theta)]^{-1}$ is the Doppler factor, with $\theta$ the viewing angle of the emission, $\Gamma$ and $\beta c$ the Lorentz factor and velocity of the outflow. Fore this estimate, we assume a homogeneous and instantaneous emission, however it can be influenced by the structure of the radiation region\cite{Rachen98, Protheroe02}. In transient sources, the short dynamical timescale of the system limits the particle escape timescale, and the diffusion of cosmic-ray particles due to scattering on magnetic waves can be considered to only intervene in the acceleration timescale. 

The energy-loss and interaction processes at play include various photohadronic and hadronic processes \cite{RachenPhD96, Dermer09}, involving interactions with a photon field of spectrum ${\rm d} n'_\gamma / {\rm d} \epsilon'$. In the case of flaring sources, the dominant interaction background to consider in the simplest case is the {\it flaring} radiation. First, interactions with steady baryon or radiation fields in the source or in the cosmic medium happen in most cases over a timescale $t \gg t_{\rm var}$, due to the larger size of the source compared to the flaring region, and due to magnetic diffusion of particles. In such a configuration, the neutrino emission will be diluted over time and can be viewed as a steady emission stemming from the quiescent source. Second, the flaring baryonic material is usually optically thinner to HE-UHE neutrino production via hadronic interactions \cite{Guepin17}. However, the case of interaction backgrounds with high opacity (for instance a dense supernova ejecta) needs to be modeled specifically.

Photohadronic processes $X\gamma$, where $X$ is a nucleon or an atomic nucleus, contribute to the production of HE neutrinos above the pion production threshold. It corresponds to the minimum proton energy $E_p' \simeq 10^{17}\,{\rm eV} \, (1\,{\rm eV} / \epsilon')$ for a typical radiation field energy $\epsilon'$. Their interaction timescales are
\begin{equation}
{t'}_{N\gamma}^{-1} = \frac{c}{2{{\gamma'}^2}} \int_{0}^{\infty} \frac{{\rm d}\epsilon'}{{\epsilon'}^2}  \frac{{\rm d}n'_\gamma}{{\rm d}\epsilon'} (\epsilon') \int_{0}^{2\gamma' \epsilon'} {\rm d}\bar{\epsilon} \,\bar{\epsilon}\, \sigma_{N\gamma}(\bar{\epsilon})\, ,
\end{equation}
where $\sigma_{N\gamma}(\bar{\epsilon})$ is the photodisintegration cross section. 
Secondary particles produced through these processes, such as electrons, pions, muons or high-energy photons, can experience energy-losses and interactions before escaping the source. Moreover, charged secondaries can be injected at shock fronts, or magnetically reconnecting regions, and be accelerated by first order or second order acceleration processes \cite{Murase12, Reynoso14, Winter14, Kawanaka15, Guepin20}. In the case of secondary particles involved in HE and UHE neutrino production, energy losses and acceleration can significantly impact the neutrino spectrum and its flavor composition.

\subsection{ Phenomenological recipes to evaluate the maximum energy and flux of HE neutrinos}\label{sec:modeling_pheno}

The competition between the aforementioned timescales allows to estimate the maximum energy of HE neutrinos. Several important assumptions usually need to be made to constrain the source properties, for instance an assumption of equipartition between the non-thermal electromagnetic energy density and the magnetic energy density to estimate the mean magnetic field. Under these assumptions, the maximum proton and neutrino energies can be estimated as a function of the transient source properties $t_{\rm var} - L_{\rm bol} - \Gamma$, as illustrated in Figure~\ref{fig:hillas}. In the same spirit as standard approaches designed to identify possible production sides for UHE cosmic rays \cite{Hillas84}, such estimates allow to identify promising categories of transient sources for the production of HE neutrino signals. In Figure~\ref{fig:hillas}, several categories of transient sources are represented, which are described further in section~\ref{sec:source_zoo}. We also give the example of PS16cgx, an object consistent with a type Ic supernova, and possibly a choked-jet or an off-axis GRB, that was identified by Pan-STARRS \cite{2016GCN..19381...1S} in a search conducted after the detection of a candidate cosmic neutrino IceCube-160427A \cite{2016GCN.19363....1B}. For deriving further properties of the source environment, the balance between the different processes at play can also be constrained using MM observations, with models dedicated to specific transient source categories.

The total non-thermal electromagnetic (EM) energy $U_\gamma$ radiated during the transient event is used as a proxy for the maximum energy channeled in HE to UHE transient neutrinos. Typically, models consider $U_p = \eta_p U_\gamma$, with $U_p$ the energy in accelerated protons and $\eta_p$ a constant. The energy channeled in neutrinos $U_\nu \simeq \eta_{\rm ch} \, f_{p\gamma} \, U_p $ depends on the efficiency of photopion production $f_{p\gamma} = \min (1, \tau_{p\gamma})$ where $\tau_{p\gamma} = t'_{\rm dyn} / t'_{p\gamma}$ is the opacity for photopion production. The factor $\eta_{\rm ch} \simeq 3/8$ depends on the channels involved in neutrino production (e.g., resonances, multi-pion production). At first approximation, the neutrino spectrum can be described by a power law between a minimum and a maximum energy, with an index $\alpha_\nu$ depending on the proton spectrum (assumed to be a power-law of index $\alpha_p$) and the photon background. For a hard radiation background, for instance a  power-law spectrum with a photon index $< 1$ or a blackbody spectrum, $\alpha_\nu = \alpha_p$ in the energy range where photopion production dominates. A softer photon spectrum or significant secondary energy losses can significantly modify the slope of the neutrino spectrum. Similar to the study of steady HE neutrino sources \cite{WaxmanBahcall98}, simple detectability criteria can be calculated using the total energy channelled in neutrinos and the properties of their spectrum, together with the sensitivities of HE to UHE neutrino experiments \cite{Rachen98, Guepin17, Guepin20, Fiorillo:2021hty}.

\subsection{Numerical methods}\label{sec:modeling_numerical}

The full treatment of HE and UHE neutrino production in astrophysical environments requires the simultaneous description of cosmic-ray propagation, acceleration, energy-losses and interactions, and the full cascade of secondary nucleons and nuclei, neutrinos and photons. Numerical simulations allow us to study these different processes in great detail. However, self-consistent simulations are challenging and computationally expensive, due to the wide range of spatial and energy scales to be considered. Moreover, the modeling of transient HE to UHE neutrino emission requires a time-dependent description of backgrounds and structures in the source environment.

The description of propagation and acceleration of cosmic rays in the vicinity of the sources requires a self-consistent treatment of the evolution of the properties of the particles coupled with the evolution of the electromagnetic fields, as the cosmic-ray energy density is not necessarily negligible when compared with the electromagnetic energy density. This coupled and non-linear description can be achieved for instance using particle-in-cell (PIC) simulations \cite{Dawson:1983zz}, but on length scales well below the typical sizes of the regions of emission inside the sources. These detailed simulations provide analytical prescriptions and initial conditions, such as acceleration timescales, particle spectra and energy densities, that can be used in larger scale simulations. Studies of HE-UHE cosmic-ray acceleration accounting for macroscopic source properties are emerging, with MHD and test-particle simulations \cite{Roh:2015cxa, Kimura:2016fjx, Kimura:2018clk, Sun:2021ods}, or fully kinetic simulations. For instance, the study of cosmic-ray acceleration in pulsar magnetospheres and winds is addressed in \cite{Guepin:2019fjb}, by making simplifying assumptions about the energy range and the energy-loss processes at play. Moreover, the addition of Monte Carlo modules to PIC simulations paves the way for self-consistent simulations \cite{Crinquand:2020ppq, 2020JPlPh..86e9016D, tristanv2}.

Simplifying assumptions can be considered to focus on specific aspects of cosmic-ray propagation and interactions in the vicinity of compact sources. Following interstellar and intergalactic propagation codes, such as GALPROP \cite{Strong:1998pw}, DRAGON \cite{Evoli:2008dv}, PICARD \cite{Kissmann:2014sia}, CRPropa \cite{CRPropa16} or SimProp \cite{SimProp17}, the propagation of cosmic ray densities at high energies can be treated by solving transport equations, and at UHE, the numerical integration of the equation of motion of single particles can be considered. These methods rely on the test-particle assumption, ignoring the feedback of cosmic rays on the electromagnetic fields. Recent work aims at unifying the simulation of cosmic-ray propagation from HE to UHE, for instance in CRPropa \cite{Merten:2017mgk}.

Approaches tackling cosmic-ray interactions and the subsequent nuclear cascades make use of Monte Carlo methods \cite{Armengaud07, Kotera09, Globus15}, while others solve coupled sets of partial differential kinetic equations \cite{Boncioli17}. Interaction cross sections and interaction products involved in HE and UHE neutrino production can be obtained from analytical formulae \cite{RachenPhD96} or numerical codes, such as \textsc{Sophia} \cite{Mucke00} for photopion production, \textsc{Talys} \cite{Talys} for photonuclear interactions and \textsc{Epos} \cite{Werner06} or \textsc{Sibyll} \cite{Riehn:2019jet} for purely hadronic interactions. Some uncertainties arise from the description of the interaction processes; for instance in the case of photonuclear interactions, the standard approach treating the nucleus as a superposition of nucleons can lead to an overestimation of the neutrino flux \cite{Morejon:2019pfu}. The implementation of cosmic-ray interactions  generally relies on an analytical description of the populations of leptons and/or the related radiation backgrounds. Therefore, the feedback of accelerated and interacting hadrons and leptons on the radiation background is only partially accounted for. The implementation of radiation backgrounds as well as interaction processes can lead to significant uncertainties on the computed neutrino fluxes \cite{AlvesBatista15, Boncioli17, AlvesBatista19}. Cosmic-ray propagation and interaction codes have been developed to study steady emissions. However, they can be adapted to study transient HE to UHE neutrino emissions by considering the evolution of interaction backgrounds in successive time-bins \cite{Globus15, Fang:2017tla, Guepin:2017abw, Decoene:2019eux}, to identify time-dependent signatures and predict coincidences or delay between emissions. Additional modules are developed, for instance in CRPropa, descriptions of Fermi acceleration, or EM cascades with thinning procedures to speed up the simulations \cite{AlvesBatista:2021mne}. Detailed studies of specific source classes can be performed, as well as large parameter-space scans are now within reach\cite{Biehl17, Heinze20}.

Another category of codes focuses on the radiation of accelerated leptons and hadrons, traditionally for the multi-wavelength modeling of active galactic nuclei spectra, for instance AM3 \cite{Gao:2016uld}, ATHE$\nu$A \cite{1995A&A...295..613M}, B13 \cite{Boettcher:2013wxa} and LeHa-Paris \cite{Cerruti:2014iwa} which are compared in \cite{Cerruti:2021hah}, and the publicly available agnpy~\cite{Nigro:2021pxy}, NAIMA~\cite{Zabalza:2015bsa} and JetSet~\cite{2020ascl.soft09001T}. These codes include synchrotron, inverse-Compton, $\gamma\gamma$ pair production or Bethe-Heitler processes, considering internal (mostly produced by the synchrotron radiation of leptons) or external radiation fields. Several of these codes can account for time-dependent emission. Generally, from the hadronic side, only nucleons are considered which is sufficient for most astrophysical environments where nucleons are dominant. At ultra-high energies the existence of appreciable fractions of other nuclear species is indicated by UHECR composition measurements~\cite{PierreAuger:2014sui}. This has led to the development of dedicated frameworks such as NeuCosmA~\cite{Hummer:2010vx, Boncioli:2016lkt, Biehl:2017zlw} and CRPropa3~\cite{AlvesBatista:2021mne,Hoerbe:2020ike}. Efforts are underway by several groups to produce coupled versions of cosmic-ray interaction and time-dependent radiative codes. 

In addition to the prediction of HE-UHE neutrino production in the vicinity of sources, the assessment of detection prospects for ground-based, sub-orbital or space-based detectors requires a good understanding of their interactions, mostly in the Earth, the ice, and the atmosphere, and the subsequent production of observable signals, such as Cherenkov light, fluorescence light, or radio waves. Several numerical packages have been developed to describe the full chain between the neutrino flux arriving on Earth, the interactions and propagation of the signals and the response of the detectors, with great precision, e.g. \cite{clsim@github,fennel2021@github,Arguelles:2021twb,NuSpaceSim}.

\section{Sources and detectability}\label{sec:modeling_sources}

Various categories of EM transients are promising sources of transient HE to UHE neutrino signals, due to their high bolometric luminosity, their powerful non-thermal emissions and/or their dense interaction backgrounds. Although they all present a common ingredient, which is a powerful plasma outflow, their structure, evolution and observational characteristics are widely different.  We provide references to recent models (section~\ref{sec:source_zoo}). Moreover, we discuss the theoretical questions that should be answered to assess their detectability, such as predictions of the neutrino energy (section~\ref{sec:source_energy}) and the variation of the neutrino signal with time (section~\ref{sec:source_time}), and we discuss the theoretical aspects required to better interpret observations --associations in particular-- through a careful modeling of theoretical uncertainties (section~\ref{sec:source_uncertainties}). We also discuss the input of data on source modeling, through non-detections or detections of multi-wavelength signals (section~\ref{sec:source_MWL}), multi-wavelength or MM coincidences (section~\ref{sec:source_coincidences}), and how these data constrain source population models (section~\ref{sec:source_populations}).

\subsection{The transient zoo at very high energies}\label{sec:source_zoo}

Several source classes can produce recurring transient emissions, such as active galactic nuclei and in particular blazars, which produce a range of flares at multiple wavelengths \cite{Gao:2016uld,Boettcher:2019gft,MAGIC:2018sak,Cerruti:2018tmc,Gao:2018mnu,Keivani:2018rnh,Reimer:2018vvw, Petropoulou:2019zqp, Oikonomou:2019pmg, Xue19, Xue:2020kuw,Paliya:2020mqm,Petropoulou:2020pqh,Rodrigues:2020fbu,oikonomou2021multimessenger,Oikonomou:2022gtz,Padovani:2022wjk,Sahakyan:2022nbz}. NS and magnetars and their sub-classes of anomalous X-ray pulsars and soft gamma-ray repeaters, as well as X-ray binaries, can also produce a range of bursts and flares \cite{Esposito:2018gvp, Zhang:2002xv, Ioka:2005er, Levinson:2001as, Anchordoqui:2002xu, Distefano:2002qw, Baerwald:2012yd, Sahakyan:2013opa}. One-time events encompass various classes of stellar-related explosions \cite{Kasliwal11}, such as novae, supernovae (SNe) and hypernovae, or super-luminous supernovae (SLSNe) \cite{Quimby11, Beall02, Arons03, Fang12, Kotera15, Guepin:2019fjb, Murase:2009pg, Fang16, Murase:2017pfe, Fang:2020bkm}, and the potentially associated low-luminosity gamma-ray bursts \cite{Liang:2006ci, Virgili09, Bromberg11, Nakar15, Salafia16, Kashiyama13, Senno:2015tsn, Boncioli:2018lrv, Fasano:2021bwq} or long-duration gamma-ray bursts (GRBs) \cite{Woosley1993, Hjorth:2011zx, Waxman:1997ti, Dermer:2000yd, Murase:2006dr, Murase:2007yt, Murase:2008mr, Murase:2013hh, Meszaros:2015krr, Globus15, Pitik:2021xhb}. 
Gravitational wave detections by the LIGO and Virgo collaborations have raised the interest for various compact object mergers, 
such as white dwarf mergers \cite{Xiao:2016man}, NS-NS mergers and the related short GRBs  \cite{Eichler89, Piro:2016jaq, Fang:2017tla, Kimura:2018vvz, Decoene:2019eux, Gottlieb:2021pzr}, NS-black hole (BH) mergers and BH-BH mergers \cite{Murase:2016etc, Kotera:2016dmp, deVries:2016ljw}, and also the disruption of stars or white dwarfs by massive or stellar mass BH \cite{FrankRees76}, namely tidal disruption events (TDEs) \cite{Komossa15, Bloom2011, Burrows2011, Murase08a, MT09, Wang11, Wang16, Senno16b, Dai16, Lunardini16, Biehl:2017hnb, Guepin:2017abw, Winter:2020ptf, Hayasaki:2019kjy, Murase:2020lnu, Liu:2020isi, Stein20}, and micro-TDEs \cite{Perets16}.  Supermassive BH mergers \cite{Yuan:2020oqg} and various classes of explosions happening in disks of active galactic nuclei \cite{Zhu:2021mqc} are also potential neutrino sources. 

\begin{figure}[!pth]
\centering
\includegraphics[width=0.65\linewidth]{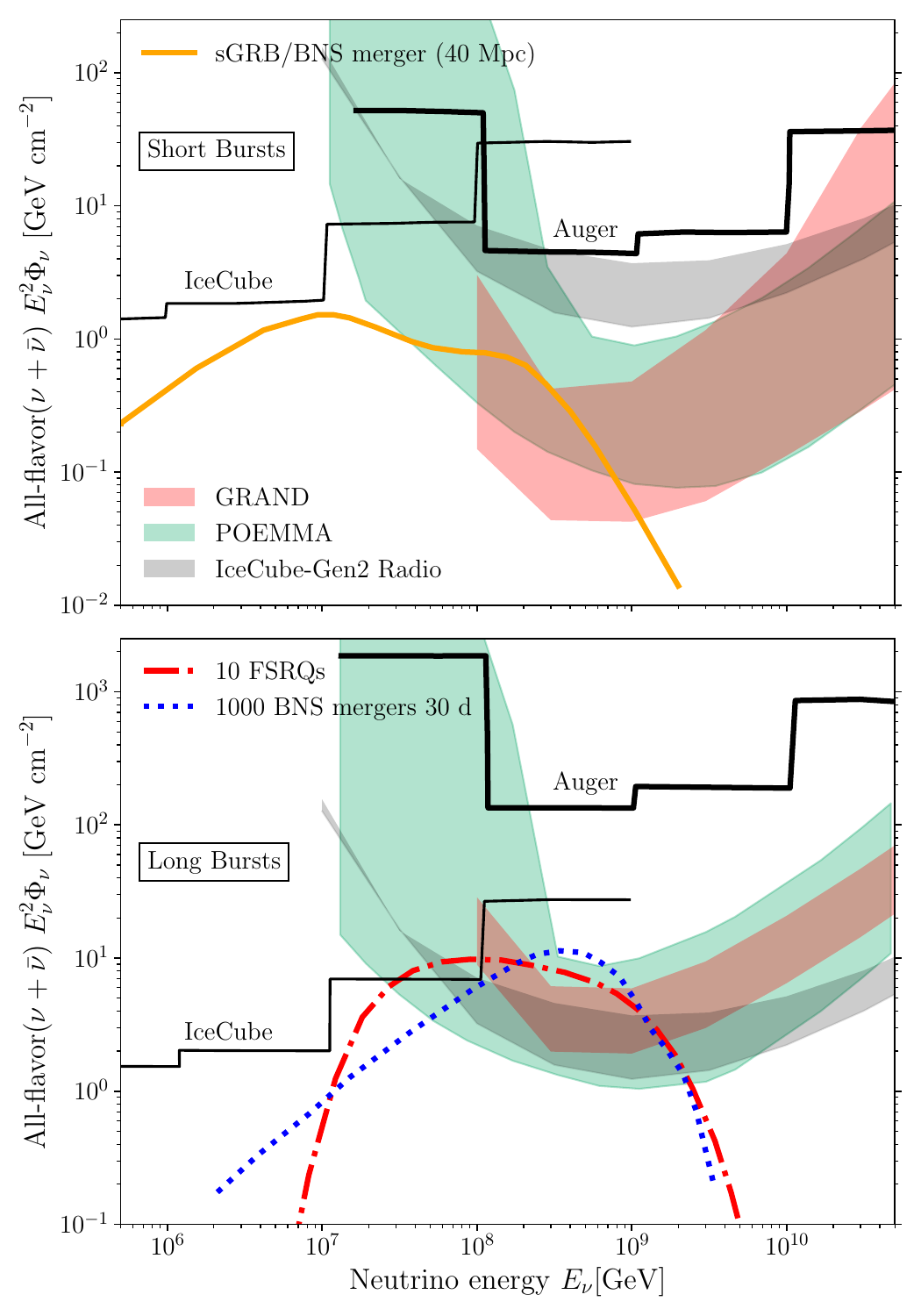}
\caption{All-flavor neutrino fluence from theoretically predicted short ($<$day,  maximum duration depending on the field of view of the instrument, top panel) and long (bottom panel) astrophysical transients and comparison to the sensitivity (instantaneous: top, and day-averaged: bottom) of existing and upcoming neutrino detectors (see text). The experimental sensitivities are per decade in energy for an assumed $E^{-2}$ neutrino spectrum. For Auger, IceCube~\cite{2017ApJ...850L..35A}, and IceCube-Gen2~\cite{2020arXiv200804323T} the sensitivity to neutrino transients at the location of GW170817A is shown. The projected sensitivity of GRAND to short transients at zenith angles $86^{\circ} \leq \theta \leq 93^{\circ}$ and long transients at declination $0^{\circ} \leq \delta \leq 45^{\circ}$ assuming a single GRAND array at latitude $42^{\circ}$ North~\cite{2020SCPMA..6319501A} is indicated with a pink band. The projected sensitivity of POEMMA to short and long transients in ToO-dual and ToO-stereo mode respectively is shown with a green shaded band~\cite{Venters:2019xwi}. The gray band indicates the IceCube-Gen2 Radio sensitivity for zenith angles $0^{\circ} \leq \theta \leq 60^{\circ}$ for short and long transients.}
\label{fig:fluence}
\end{figure}

Figure~\ref{fig:fluence} presents theoretically expected neutrino fluences for a number of promising astrophysical source classes. Overlayed are the projected all-flavor neutrino fluence sensitivity limits for upcoming instruments, i.e, the Feldman-Cousins upper limit\cite{Feldman:1997qc} per decade in energy at 90\% C.L., assuming a power-law all-flavor neutrino spectrum $E_\nu^{-2}$, for no candidate events and null background. For Auger, IceCube~\cite{2017ApJ...850L..35A}, and IceCube-Gen2~\cite{2020arXiv200804323T} the sensitivity to neutrino transients at the location of GW170817A is shown. 

The upper panel (labeled "short bursts") presents the instantaneous sensitivity, for a short transient source located in the instrument field of view. The orange line corresponds to a theoretically expected neutrino fluence of a short GRB associated with a binary neutron star merger located at 40~Mpc in the case of on-axis viewing~\cite{Kimura:2017kan}. The lower panel ("long bursts") corresponds to the daily averaged sensitivity: for each declination, the experimental effective area is averaged over the right ascensions. The detectability of bursts with duration longer than the time for a source to remain in the field of view of narrow instruments should be assessed with these daily averaged sensitivities. The long transients shown are: {\it(i)} Stacked ten-year neutrino fluence from ten flat spectrum radio quasars (FSRQs), weighted according to the gamma-ray activity. The neutrino emission is dominated by the gamma-ray flaring states, based on the infrared torus model of~\cite{oikonomou2021multimessenger}. {\it (ii)} Stacked neutrino signal from 1000 NS-NS-merger events~\cite{Fang:2017tla}. Several of the theoretical models will be tested by the upcoming neutrino facilities. 

For GRAND, the pink band encompasses zenith angles $86^{\circ} \leq \theta \leq 93^{\circ}$ for short transients and declination $0^{\circ} \leq \delta \leq 45^{\circ}$ for long transients, assuming a single GRAND array at latitude $42^{\circ}$ North~\cite{2020SCPMA..6319501A}.
The green band indicates the sensitivity of POEMMA to short and long transients in ToO-dual and ToO-stereo mode respectively~\cite{Venters:2019xwi}. The gray band represents the IceCube-Gen2 Radio sensitivity, using the same definition as for the other experiments, for zenith angles $0^{\circ} \leq \theta \leq 60^{\circ}$. As IceCube-Gen2 Radio has a homogeneous effective area over all right ascensions, the instantaneous and daily averaged sensitivities are equal.

\subsection{At what energies are neutrinos typically produced?}\label{sec:source_energy}

As can be seen in Figure~\ref{fig:fluence}, the typical energy of neutrinos produced by the various source classes has a decisive impact on their detectability with a given observatory. At first approximation, transient neutrino production is described by one-zone models that focus on one-time interval around the peak of the EM emission. In these cases, as illustrated in Figure~\ref{fig:hillas}, the maximum energy of neutrinos depends on cosmic-ray acceleration, and energy losses of primaries and secondaries. Long duration transients such as SLSNe, blazar flares or TDEs, which are marginally subject to energy losses (such as synchrotron radiation) for primary hadrons or secondary particles, can produce neutrinos up to the UHE range. Short duration luminous transients, such as millisecond magnetars, magnetar bursts and flares, NS-NS mergers or GRBs, are also good candidates for cosmic-ray acceleration to EeV energies and above \cite{Fang:2015xhg, Fang:2017tla}; however, due to their short duration and the related small radiation zone, their intense magnetic fields and radiation backgrounds can in specific cases significantly limit the maximum energies of transient neutrinos to the HE-VHE range or below. Dense radiation or hadronic backgrounds can also significantly reduce the efficiency of the acceleration processes.

\subsection{How does the neutrino signal vary with time?}\label{sec:source_time}

The hints of spatial association of HE neutrinos with transient sources, see section~\ref{sec:source_coincidences} for concrete examples, combined with their detection after the peak EM emission, motivates an accurate description of transient neutrino spectra and lightcurves. This requires time-dependent modeling of acceleration and energy-loss processes, which is of particular interest for specific transient sources with intense MM follow-up \cite{Stein20}. For various source classes, the peak fluxes and peak energies of transient photons and neutrinos can occur at different times, as acceleration and energy-loss processes evolve through the duration of the transient. For instance, models of non-relativistic shock-powered transients show a delay between the peak flux of HE neutrinos and the time when they are produced at a maximum energy \cite{Fang:2020bkm}, which can impact detectability for HE to UHE neutrino detectors. Moreover, secondary energy losses and interactions can significantly suppress HE to UHE neutrino production during the early phase of the transient EM or MM emission, with a neutrino energy and flux peaking at later times, for instance in the case of NS mergers \cite{Fang:2017tla, Decoene:2019eux}. Similarly, the interactions of VHE photons, for instance through pair production processes, can suppress a part of the transient multi-wavelength emission and induce delays between the emissions of neutrinos and VHE photons. In some scenarios with high opacities, HE to UHE transient neutrinos can appear as precursors of the VHE EM emission, for instance for models of chocked GRBs \cite{Meszaros:2001ms, Senno:2015tsn}, despite potential strong flux and maximum energy suppression due to meson cooling \cite{Horiuchi:2007xi}. 

\subsection{Model uncertainties}\label{sec:source_uncertainties}

Several sources of uncertainties challenge the modeling of transient neutrino signals. Different emission regions in winds or jets can usually explain the EM emission of a given source, with significantly different parameters. The magnetic field, Lorentz factor, and the parameters characterizing the radiation and hadronic backgrounds cannot systematically be inferred from observations and can be degenerate, which directly impacts the acceleration and the energy losses of primary and secondary particles involved in the neutrino production, and the efficiency of neutrino production. The luminosity of the cosmic rays accelerated inside the source is loosely constrained, as well as their composition, which can strongly impact the neutrino spectrum due to the photodisintegration of heavy nuclei, e.g. through giant dipole resonance, and the production of secondary nucleons and nuclei. The uncertainties on the interaction backgrounds also impact the opacity of the source to EM radiation, and thus the estimate of the energy released during the flare. Recent studies compare different emission models, e.g. \cite{Keivani:2018rnh,Oikonomou:2019djc,Pitik:2021xhb} or systematically explore a fraction of the parameter space, e.g. \cite{Gao:2016uld,MAGIC:2018sak,Cerruti:2018tmc,Rodrigues:2018tku,Reimer:2018vvw,Heinze20}.

\subsection{Learning from multi-wavelength data}\label{sec:source_MWL}

Precise multi-wavelength and time-dependent data are decisive for distinguishing between transients, such as jetted TDEs and AGN, or non-jetted TDEs and SNe; and draw notable comparisons between the various source models. The amount of data available and coverage of the EM spectrum differs between transient categories, depending on the source locations and populations (Galactic, extra-galactic) and the possibility of conducting multi-wavelength follow-up campaigns. For instance, the small number of detections of short GRBs and NS-NS mergers remains a limitation to characterize their properties and their population. Specific EM constraints of the cosmic-ray component in sources are starting to emerge. Some VHE EM emissions can be interpreted hadronic emissions, however they can often also be explained by leptonic emissions. The absence of detection of VHE EM emission can be a marker of interaction backgrounds with a high optical depth and thus intense HE-UHE neutrino production \cite{Murase:2015xka}, accompanied by EM cascades of VHE photons. In addition, EM cascades related to hadronic processes can provide strong constraints for specific source classes, with dedicated observations in the MeV-GeV energy range \cite{Gao:2018mnu, Keivani:2018rnh}. However, intergalactic magnetic fields can affect EM cascades and thus the arrival times of gamma rays \cite{Halzen:2018iak, AlvesBatista:2020oio}. This important effect can significantly impact correlation studies.

\subsection{Learning from coincidences}\label{sec:source_coincidences}

The first hints of associations between EM emissions and HE neutrinos detected by IceCube, from blazar flares\cite{IceCube:2018dnn, IceCube:2018cha, 2019ATel12967....1T, 2020A&A...640L...4G} TXS 0506+056 + IC170922A, PKS 1502+106 + IC190730A, 3HSP J095507.9+355101 + IC200107A, and TDE\cite{Stein20} AT2019dsg + IC191001A, or probable TDEs\cite{Reusch:2021ztx} AT2019fdr + IC200530A and \cite{2021arXiv211109391V} AT2019aalc + IC191119A further motivate the on-going modeling effort for predicting time-dependent MM and neutrino emissions from these sources. These associations, if physical, imply extremely high proton luminosity, exceeding the Eddington luminosity of the sources, and very high photo-pion induced EM cascade flux in the keV-GeV energy range as a result. The cascade flux, and thus the proton luminosity can be constrained by multi-wavelength observations.

Many spatial and temporal correlation studies between detected HE neutrinos and individual or population of sources are conducted. For instance, hints for correlations have been found between intermediate and high-energy peaked blazars and HE neutrinos \cite{Giommi:2020hbx} and radio-selected blazars~\cite{Plavin:2020mkf}; these associations being individually compatible with statistical fluctuations. Among other examples, time-dependent searches are performed by IceCube for increased X-ray activity of magnetars \cite{Ghadimi:2021btz}, using the McGill Online Magnetar Catalog \cite{Olausen:2013bpa}, X-ray binaries \cite{IceCube:2021vlt}, using their X-ray lightcurves, gravitational wave detections of LIGO/Virgo~\cite{IceCube:2021ads}, supernovae~\cite{IceCube:2021oiv}, GRBs~\cite{IceCube:2016ipa, IceCube:2021qyj}, blazar flares~\cite{IceCube:2021oqh}, minute-scale transients in general~\cite{IceCube:2018omy}. General searches for associations with optical transients and resulting limits have been presented in~\cite{Necker:2022tae,Stein:2022rvc}.

Associations are expected to significantly increase in the next decades, with the multiplication of observations of transient sources, and in particular observations of compact object mergers \cite{2019ARNPS..69..477M}. As an example, in the past three years, the Zwicky Transient Facility has discovered $> 1500$ core-collapse supernovae, $> 100$ hypernovae and $> 30$ tidal disruption events (among other transients). The modeling of possible coincidences will require a precise description of spatial properties, to correctly assess emissions directed towards the observer; for instance, in the case of GW170817, to correctly predict coincident photons and neutrinos that are not emitted in the direction of the relativistic jet (off-axis emissions) \cite{Biehl:2017qen, Ahlers:2019fwz}.

Finally, in coincidence studies, a careful consideration should be given to HE neutrino backgrounds. The contribution of all HE neutrino production regions along the line of sight from the Earth, not necessarily located in the vicinity of transient sources \cite{Fang:2016amf, Fang:2017zjf, Hussain:2021dqp}, should be accounted for. This irreducible astrophysical neutrino background must be accounted for when calculating the significance of the detection of HE neutrinos from transient sources. The impact of this astrophysical background on the significance of observed associations depends on the precision of temporal and spatial coincidences, instrument properties, neutrino emitting source population, and the energy of the neutrinos.  

\subsection{Constraints on source populations}\label{sec:source_populations}

The contribution of observed GRBs to the diffuse HE neutrino flux detected by IceCube is strongly constrained by time-dependent searches \cite{IceCube:2016ipa, Albert:2016eyr, ANTARES:2020dpd}. The contributions of known source populations to the total cosmic neutrino flux detected by IceCube, constrained through stacking analyses \cite{Yuan:2019ucv, Bartos:2021tok}, are the following: prompt phase of GRBs ($<1\%$)~\cite{IceCube:2016ipa}, blazars ($<17\%$)~\cite{Huber:2019lrm}, non-jetted TDEs ($<26\%$) and jetted TDEs ($<1\%$)~\cite{Stein:2019ivm}, ultraluminous infrared galaxies ($<10\%$)~\cite{ICECUBE:2021edr}, Galactic sources ($<14\%$)~\cite{2017ApJ...849...67A}, 
SNe IIn ($<55.2\%$), SNe IIP ($<79.6\%$) and SNe Ibc ($<28.6\%$)~\cite{IceCube:2021oiv}, microquasars ($<7.3\%$)~\cite{IceCube:2021vlt}, galaxy clusters out to redshift 2 (<5\%)~\cite{IceCube:2021abh}, and radio galaxies ($<30\%$)~\cite{Zhou:2021rhl}. 
Recently, a $2.6\sigma$ excess of neutrinos from the directions of non-jetted AGN weighted by their soft X-ray flux has been reported~\cite{IceCube:2021pgw}. If this is a genuine signal, it implies that 27\% to 100\% of neutrinos at 100 TeV come from non-jetted AGN.

Uncertainties related to the relation between the electromagnetic flux at a specific wavelength and the neutrino flux of the sources as well as the exact shape of the emerging neutrino spectrum significantly impact these estimates. 

\begin{figure}[ht]
\centering
\includegraphics[width=0.65\linewidth]{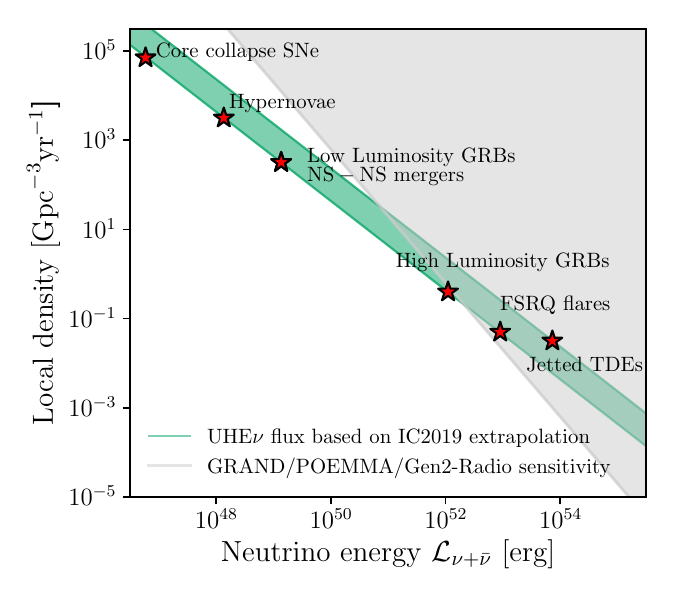}
\caption{Comparison of the local density and implied neutrino energy output of transient extragalactic neutrino sources to the sensitivity of upcoming UHE neutrino detectors. The green band indicates the neutrino energy output / and local number density range that is compatible with the total observed diffuse neutrino flux in the IceCube energy range extrapolated to $10^{18}$~eV using the best-fit parameters of~\cite{IceCube:2021uhz} for sources whose redshift evolution ranges from negative to positive and equal to the redshift evolution of the star formation rate. The grey shaded area indicates the parameter space that will be constrained by upcoming detectors in the absence of detection of point sources.}

\label{fig:rate_density}
\end{figure}

A luminosity dependent lower limit on the number density of neutrino sources can be derived from the absence of significant small scale clustering (multiplets) and neutrino point sources in the IceCube data. Their absence disfavours rare and very bright astrophysical sources as the sources of the bulk of the neutrino flux in the IceCube energy range~\cite{Lipari:2008zf, Ahlers:2014ioa, 2016PhRvD..94j3006M, Murase:2018iyl,Ackermann:2019ows,Capel:2020txc}. A similar constraint will be possible to derive at ultra-high energies with planned neutrino facilities if a UHE neutrino flux is detected. The expected capabilities of GRAND, IceCube-Gen2-Radio, and POEMMA assuming their planned limiting sensitivity $F_{\rm lim} = 0.1$GeV/cm$^2$ at $10^{18}$~eV are shown in Figure~\ref{fig:rate_density}. The neutrino flux at energy exceeding $10^{16}$~eV has been estimated extrapolating the parameters that best describe the IceCube diffuse flux at lower energies, as derived by~\cite{IceCube:2021uhz}. The width of the green band corresponds to the uncertainty introduced by the unknown redshift evolution of neutrino sources for which we considered the range considered by~\cite{Palladino_2020}. The assumed rates of the transient events considered are according to the observed rates quoted in Table 2 of~\cite{Murase:2018utn}, except for NS-NS mergers for which we have used the updated estimate of\cite{2021ApJ...913L...7A}. For the observed duty cycle of FSRQ flares we have taken the mean of the estimates of\cite{Fermi-LAT:2019yla} and \cite{Fermi-LAT:2021ykq}. In some instances, most demonstrably jetted TDEs, the power in cosmic rays must exceed the EM power of the objects so as for them to account for the bulk of the high-energy neutrino flux. This calculation is conservative in the sense that it doesn't take into account the possibility of an additional UHE neutrino emitting source population above $10^{16}$~eV, which would lead to enhanced neutrino flux in the considered energy range with respect to what is expected based simply on an extrapolation of the TeV-PeV astrophysical neutrino flux. For an independent calculation assuming different benchmark scenarios see~\cite{Fiorillo:2022ijt}.

The non-detection of joint gravitational waves and neutrino events~\cite{2019ApJ...870..134A} also starts to set constraints on the population density of powerful and high-rate transients capable of producing copious gravitational wave signals, such as short gamma-ray bursts~\cite{IceCube:2020xks}.

\section{Instruments to detect and identify high-energy neutrino transients}

Further constraints on source modeling and the actual identification of neutrino transients will be achieved by adequate and powerful neutrino detectors, combined with EM follow-up telescopes and gravitational wave detectors via efficient networks. We discuss the performances of such current and up-coming instruments in the following, focusing on neutrino and EM observatories. Because of the pointing capabilities of EM instruments compared to neutrino and gravitational-wave detectors, a joint neutrino-gravitational-wave detection would be detected by correlating neutrino and EM signals on the one hand and with EM-gravitational-wave campaigns on the other. Various reviews cover the latter subject~\cite{miller19,Bailes21}, hence we concentrate on the former. 

\subsection{Neutrino detectors}

Neutrino detection requires large target volumes because of their low interaction probability with matter. At very high energies (up to $\sim 10$\,PeV), large volumes of ice (for IceCube) or water (for ANTARES) have been instrumented, in order to detect the Cerenkov light emitted by up-going neutrinos crossing the Earth or down-going neutrinos arriving  at small zenith angles. The successful detection of PeV neutrinos with IceCube~\cite{2014PhRvL.113j1101A, 2015PhRvL.115h1102A} has triggered several international efforts to build larger detectors following the same principles~\cite{2021JKPS...78..864R}: KM3NeT-ARCA~\cite{2016JPhG...43h4001A}, Baikal-GVD~\cite{2021arXiv210714303A} and P-ONE~\cite{P-ONE_2020} deep underwater, and IceCube-Gen2~\cite{2020arXiv200804323T} in ice. Note that several of the projected instruments are based on detection techniques that have not been fully tested. For example, the success of BEACON, GRAND, POEMMA, TAMBO, or Trinity rely on reconstructing air-showers in Earth grazing configurations, or from above the atmosphere. GRAND also requires to perform efficient self-triggering on radio signals, which still needs to be  validated, although promising preliminary studies exist \cite{CHARRIER201915}.

At higher energies, the fluxes are lower, and more cost-effective detection techniques have to be utilized to instrument even larger volumes. Earth-skimming tau-neutrinos can interact with the Earth and produce tau leptons which can decay and lead to air-showers~\cite{Fargion:2000iz}. These air-showers can be observed via their particle (Auger~\cite{Aab_2019}, TAMBO~\cite{2020arXiv200206475R}) their air-fluorescence or beamed optical Cerenkov (Auger, POEMMA~\cite{2021cosp...43E1367O}, Trinity~\cite{2021arXiv210802751W}, Ashra-NTA~\cite{2014arXiv1408.6244S}) or radio (GRAND~\cite{2020SCPMA..6319501A}, BEACON~\cite{2019BAAS...51g.191W}) emissions. Radio-detection techniques can also be used to search for the coherent, impulsive radio signals that are emitted
by electromagnetic particle cascades induced by neutrinos interacting with ice~\cite{2017nacs.book..217C}, in ice (ARA~\cite{2019arXiv190711125A}, ARIANNA~\cite{2020arXiv200409841A}, RNO-G~\cite{2021JInst..16P3025A}, IceCube-Gen2 Radio~\cite{2020arXiv200804323T}) or with balloon experiments (ANITA~\cite{2019ICRC...36..867D}, PUEO~\cite{2021JInst..16P8035A}).

In light of the previous sections, future neutrino observatories should meet specific requirements in order to optimize the probability of detecting HE and UHE transients. A fluence sensitivity improvement of one (two) order(s) of magnitude in the PeV (EeV) range, should be aimed for, in order to reach the levels of the fluxes of the most powerful transient source categories. A maximal coverage of the energy range from PeV to EeV is desirable, with an energy resolution of $\Delta E/E\lesssim 0.5$ to distinguish between different source spectra. For efficient follow-ups, coincident searches, and source identifications sub-degree angular resolution will be necessary, as well as wide instantaneous (or daily, or monthly --depending on the type of transient) field of view (FoV), and a multi-instrument network that enables full sky coverage. Finally, being able to send and receive alerts is a key requirement.

Current and future neutrino detectors are listed in Table~\ref{table:nu_instr_char} together with their timeline and their indicative characteristics. To give an idea of the span of the energy range of the experiment, we quote the minimum energy --for most experiments this either corresponds to an operational energy, or is defined as the energy above which the sensitivity is competitive with the IceCube limits (above PeV) over the lifetime of the project-- and the ``peak'' energy, where the instrumental sensitivity is best. Projected differential sensitivity limits to the diffuse neutrino flux are quoted at this peak energy. Although fluence sensitivities would be better suited to assess transient source detectability, they are less convenient to quote due to dependencies on declination and time integration. The sensitivities listed here can serve for comparison between instruments. We caution that for many of the projected instruments, these numbers are preliminary, based on simulations and/or extrapolations from existing measurements\footnote{For example, for Baikal-GVD, the predicted sensitivity is for a full detector of 112 strings, extrapolated from the current instrument of 64 strings}. For some experiments, we quote the existing upper limits in brackets, and the measured flux for IceCube. The instantaneous field of view (iFoV) and the daily sky coverage (dFoV) are given as percentage of visible sky (the observable band is also quoted for some experiments in square degrees in parenthesis). For the very high-energy instruments ($<10\,$PeV), geometrical FoVs are listed, that do not fold in any cut on the acceptance of the instrument\footnote{For example, with a 90\% cut on the acceptance, the FoV of P-ONE would drop to 22\% (instantaneous) and 24\% (daily average).}. For the UHE instruments, a cut at 95\% acceptance was taken into account for the FoV calculations of ANITA, PUEO, ARA, POEMMA fluorescence, Auger, GRAND, and Trinity. For proposed instruments, the FoV could change depending on the final chosen site. The FoV and angular resolution also depend on the neutrino energy and on the nature of the event. In this table, the best performances are quoted, usually around the peak energy. For very high-energy instruments, when the data is available, a distinction is made for up-going muons tracks (denoted `up') and cascade events (marked in parenthesis), or for combined (`up+casc.') events when specified\footnote{Muon tracks are elongated signals from muon neutrino interactions in or near the detector. Electron or tau neutrinos rather deposit most of the neutrino energy in a small region, resulting in nearly spherical `cascade' events, from which it is more difficult to infer the arrival direction.}.

The table indicates that thanks to their complementary strengths and locations, the projected experiments have a good potential of fulfilling most of the requirements listed above for successful transient neutrino detection. The planned PeV energy-range telescopes observe different parts of the sky and their different techniques will enable to precisely study systematic effects. Moreover, their characteristics indicate that they will have complementary strengths: IceCube-Gen2 will accumulate the largest statistics, while KM3Net will achieve a an angular resolution more than 3 times better, thanks the less diffusive water-medium, compared to ice. For UHE instruments, in-ice radio, optical and particle detectors reach lower energy thresholds, which ensures a continuous energy coverage from the PeV range. In-ice radio experiments also provide a wide instantaneous FoV, ideal for transient monitoring. For these instruments however, although the angular resolution on the arrival angle of a signal is of sub-degree order, the reconstruction of the direction of the incident neutrino is more complex, as it is dominated by the resolution on the orientation of the signal polarization~\cite{Kim:2019BK,Barwick:2021wU}. Hence, the angular resolution is expected to be $>1^\circ$. On the other hand, radio-air shower detectors and some optical instruments will naturally provide angular resolutions of $<1^\circ$. The combination required for neutrino astronomy as discussed above (a sensitivity improvement of 2 orders of magnitude, sub-degree angular resolution and wide field of view) is projected for two instruments: GRAND and Ashra-NTA.

We have not quoted energy resolutions in Table~\ref{table:nu_instr_char}. Reaching good energy resolutions might be challenging for all detectors observing muon and tau neutrinos. Although the energy of the air-shower resulting from the lepton decay could be resolved within $10\%$, the uncertainty on the energy rate received by the lepton is the limiting factor ($\Delta E/E\sim 0.3-0.5$)~\cite{2014JInst...9P3009A,2021arXiv210702604A}, but this could suffice to differentiate between source spectra. 

The last column of Table~\ref{table:nu_instr_char} lists the neutrino alert programs that detectors have or are planning to set up. Current networks and alert systems are indicated in roman style. Most instruments (marked as ``planned'') are planning to send alerts. When network names are specified in italics, the corresponding instruments have started to implement alert programs or are working on extensions of existing structures. For some instruments like ARA, ANITA and PUEO, implementing an alert system requires fundamental changes in data handling, hence there are no immediate plans to be part of an alert system.

\subsection{The power of multi-messenger networks}

The IceCube observatory has provided us with a first view of the neutrino sky. From the absence of clearly detected point sources and of small scale clustering, we have learnt that the sources of high-energy neutrinos are relatively abundant as discussed in Sec~\ref{sec:modeling_sources}. In light of the fact that the sources are relatively numerous and unlikely to lead to clear detection with neutrinos alone, a promising method to identify neutrino sources is by association of a neutrino with an electromagnetic signal accompanying a transient event responsible for its generation. The required temporal coincidence in addition to a spatial association increases the sensitivity and the significance of a potential discovery. 

The power of such EM+neutrino associations is harnessed by several ongoing programs. These programs are based on two main observation strategies: 
A) EM network follow-up on neutrino alerts~\cite{mathieu:tel-01237558,2017APh....92...30A}, and
B) neutrino telescopes following up on EM transient alerts~\cite{2021arXiv210709551P, Dornic:2021kko}. 

As part of A), the IceCube realtime alert system~\cite{2017APh....92...30A} rapidly communicates the arrival direction of neutrinos with a high-probability of being astrophysical, via a Gamma-Ray Coordinates Network (GCN) alert\cite{GCN}, to allow for follow-up with steerable telescopes. IceCube generates approximately ten "Gold" alerts per year\cite{Blaufuss:2019fgv}. Such neutrinos have $\sim 50\%$ chance of being astrophysical as opposed to atmospheric, meaning that about half of these alerts point to truly astrophysical neutrinos. Additional alert classifications exist with a larger fraction of background neutrinos. X-ray, optical, and gamma-ray instruments typically have a number of hours, or pointings allocated to such follow-up,  as part of their target of opportunity (ToO) programs (Table~\ref{table:instr_char}). Non-steerable detectors including Auger, ANTARES perform archival searches for similar emission~\cite{10.3389/fspas.2019.00024,Dornic:2021kko}. 

In addition to bilateral collaboration programs, dedicated networks focus on specific types of associations, as part of observation strategy B). Since 2005 the SuperNova Early Warning system (SNEWs) has been running with the aim to alert the astronomical community at the time of next galactic supernova, by requiring an online coincidence among several world-wide distributed neutrino detectors~\cite{Vigorito_2011}. The Astrophysical Multimessenger Observatory Network (AMON) is a cyberinfrastructure network which connects ANTARES, Auger, FACT, Fermi-GBM, Fermi-LAT, H.E.S.S., HAWC, IceCube, LCGOT, LIGO-Virgo, LMT, MASTER, Swift, VERITAS, ZTF and focuses on coincidences among two or more of these instruments in what would be subthreshold events in one individual detector~\cite{AyalaSolares:2019iiy,619f9a987a014aa5bf0214e60b7aa160}. In addition, the AMON infrastructure is used as a pass-through for the aforementioned IceCube high significance alerts. Other instruments are at the time of this publication in the process of joining AMON. The increasing rate of astrophysical transients which are detected with astronomical surveys, particularly optical ones, has lead to alternate approaches complementary to those of MM networks. The Alert Management, Photometry and Evaluation of Lightcurves (AMPEL) system~\cite{2019A&A...631A.147N}, processes astronomical data according to user defined criteria, in order to produce sub-samples of events, relevant to a particular MM search. AMPEL is particularly useful for processing the data acquired with surveys such as ZTF. 

\subsection{Electromagnetic follow-up instruments}

The entire EM spectrum is today covered for coincident searches, with a combination of wide field survey instruments (e.g., LHAASO~\cite{bai2019large}, HAWC~\cite{doi:10.1063/1.2751958}, ASAS-SN~\cite{2014ApJ...788...48S}, ATLAS~\cite{2018PASP..130f4505T}, PanSTARRS~\cite{2016arXiv161205560C}, ZTF~\cite{Bellm_2018}, Vera Rubin Observatory~\cite{2019ApJ...873..111I}), high angular resolution pointing telescopes (e.g., CTA~\cite{maier2019performance}, H.E.S.S.~\cite{hess},  Kanata/HONIR~\cite{2021PASJ...73...25M,2021GCN.29820....1S}, MAGIC~\cite{Aleksi__2016}, MASTER~\cite{2010AdAst2010E..30L}, VERITAS~\cite{park2015performance}, XMM-Newton~\cite{XMM}, Athena~\cite{2013arXiv1306.2307N}, TAROT~\cite{boer:hal-01511688}), multi-platform satellites (e.g., {\it Fermi}~\cite{Garrappa:2021vT}, INTEGRAL~\cite{Ferrigno_2021}, {\it Swift}~\cite{2004ApJ...611.1005G}, SVOM~\cite{2016arXiv161006892W}) large field-of-view imaging radio facilities (e.g., VLA~\cite{2018SPIE10700E..1OS}, MWA~\cite{2009IEEEP..97.1497L}, MeerKAT~\cite{Jonas:2018Jr}, SKA~\cite{2017arXiv171206950A}), and spectroscopic programs to measure the source redshift (e.g., GEMINI GMOS~\cite{2004PASP..116..425H}, GTC OSIRIS~\cite{10.1117/12.460913}, Keck LRIS~\cite{1995PASP..107..375O,2010SPIE.7735E..0RR}, VLT X-shooter~\cite{2011A&A...536A.105V}). Table~\ref{table:instr_char} compiles indicative experimental characteristics --detailed below-- of a non-exhaustive list of actual or potential neutrino follow-up EM instruments. 
The majority of the listed experiments have a dedicated ToO follow-up program, and all of them, except for ATLAS, have participated in the follow-up of at least one neutrino alert. 

Although they are not listed in Table~\ref{table:instr_char}, future gamma-ray observatories in the MeV range such as AMEGO \cite{2019BAAS...51g.245M}
with capabilities beyond INTEGRAL and Fermi-GBM, will fill an important gap and enable new types of studies. They will extend the flux measurements of HE gamma-ray sources to lower energies, connecting with X-ray observations, in an energy range where gamma-rays are not absorbed by the extragalactic background light. 

\subsubsection{Field of view}
Coincident observations of a neutrino and a transient counterpart require that the field of view (FoV) of the telescope be comparable to the angular resolution of the neutrino detector. The FoV depends on the energy/wavelength and on the observation mode of the facilities. We quote the FoV at or in a range around the best sensitivity of the instruments.\footnote{For ASAS-SN, combining the 4 sites, the FoV is of 360 square degrees. However, when responding to a ToO, the best single unit of 4 cameras is triggered, leading to a narrower FoV of 72 square degrees~\cite{2014ApJ...788...48S,2017PASP..129j4502K}. The figure for this latter observation mode is quoted.}

\subsubsection{Slew speed and survey speed}
The efficiency of an early follow-up program further depends upon the fast response and slew speed of the telescopes after receiving the alert. The indicated response delay corresponds to the time between the instant when a neutrino or ToO alert is received by an instrument, and when it starts the follow up. It can range from seconds or minutes for gamma-ray experiments in particular, to days for wide field instruments that require Earth rotation to reach the required field of view in the sky. For spectrographs, the slew rate is given in terms of typical time $t_{\rm slew}$ between the pointing of two objects in the visible sky, as ``obj.\,/\,$t_{\rm slew}$''. HAWC and LHAASO do not have a response delay, as they do not point to particular directions of the sky. They can follow up on any source that falls in the visible sky within a 24-hour rotation of the Earth, but their exposure depends on the zenith angle, such that some short transients in some specific position of the sky might not be detected.
For survey-type instruments, transient detection probability also crucially depends on survey speed or cadence (quoted in brakets for normal survey mode), i.e., the mean time between visits to the same piece of sky. 

\subsubsection{Differential sensitivity limits}\label{section:EM_diff_sens}

For high-energy experiments, the differential sensitivity limit quoted in Table~\ref{table:instr_char} corresponds to the minimum flux needed by an instrument to obtain a   5-standard-deviation detection (4$\sigma$ for XMM-Newton) of a point-like source, for the indicated observation time, as available in the literature. The relative scalings can differ of a factor of order unity between the experiments, due to different energy binning, detection criteria and calculation methods. For optical telescopes, a typical depth in magnitude over the different bands, and for radio observatories, the sensitivity for a characteristic frequency (e.g., the dipole resonance frequency) and bandwidth, with the instrument in its best configuration, are quoted. For ASAS-SN and ZTF, the sensitivity indicated is for dedicated ToO images with longer exposure time compared to the normal survey.

\subsubsection{Angular resolution}

The best or typical (in the given energy range) angular resolutions at FWHM, are indicated for each instrument. For the four listed optical spectrographs, the angular size of the pixels has been referenced. 

\subsubsection{Follow-up programs}

The final column gathers information on the neutrino follow-up programs of these instruments. Most high-energy instruments have dedicated neutrino or more generally ToO follow-up time. The indicated neutrino follow up rates (in hours per year) should be taken with great caution, as these number change every year in the internal proposal evaluation processes. Besides, the time allocated to a certain program does not necessarily reflect how many observations are actually being performed. Each collaboration has different priorities and complementary ways of running follow-up programs. For example H.E.S.S. aims at deep and long-term follow-ups of a few selected events while MAGIC covers as many neutrinos as possible but with short exposures~\cite{Acciari:2021YA}.

In the same column, in brakets, we also indicate the number or percentage of neutrino alerts that have been followed up. The alerts are either published GCN alerts by IceCube or ANTARES, or IceCube Gold alerts. The fraction of followed alerts depends on the accessible sky and reflects the internal strategies and cuts set by each program. For instance, thanks to the numerous sites positioned around the globe, MASTER follows 99\% of the published GCN neutrino alerts. ASAS-SN, TAROT and ZTF follow most accessible IceCube neutrino alerts, that are not behind the sun or below the horizon. 
Interestingly, these do not represent a large fraction of the observing time for these telescopes (< 5\%). On the other hand, MWA prioritizes immediate triggering rather than late follow up (if an alert is not issued in a portion of sky that is not currently in the field of view of MWA, it will not follow up later, once the Earth has rotated). Its large FoV however provides a reasonable chance to get late-time follow-up ``for free'' enabling the array to derive interesting constraints on the radio emissions in the direction of ANTARES neutrinos~\cite{Croft_2016}. Instruments such as {\it Fermi}-LAT and HAWC have a wide FoV that enables one to get information about any position in the sky almost continuously, resulting in almost 100\% of followed alerts. Very recent or planned instruments such as LHAASO, CTA or Rubin Observatory are elaborating their follow-up programs, and the allocation times will be adjusted over the next few years. 

For some instruments which have only followed up a limited number of neutrino alerts, we give in the same column (in italics) specific examples of sources or events that were studied: TXS 0506+56~\cite{2017ATel10840....1C,Paiano_2018,Keivani17,2017ATel10861....1T}, IC190331A~\cite{Krau__2020}, SN PTF12csy~\cite{Aartsen_2015}, and few ANTARES events for VLA~\cite{2015ATel.8034....1T,2015ATel.7999....1H}. VLT spectrographs X-Shooter and FORS2 as well as GTC spectrograph OSIRIS have also participated in the spectroscopy of a sample of blazars that are neutrino source candidates~\cite{Paiano_2021}. For XMM-Newton, we also quote a few follow-up examples. Its relatively slow slewing speed compared to Swift and the limits on the solar panel orientation with respect to the Sun (which reduces significantly all-sky visibility), make it not ideal for fast repointing on ToOs. As a consequence, XMM-Newton has only a limited history of neutrino follow ups, e.g., the ``Kloppo'' PeV IceCube event~\cite{Aartsen_2016}, or PKS 1502+106~\cite{2019ATel12996....1K,Abdo_2010,Rodrigues:2020fbu,oikonomou2021multimessenger}.

\subsubsection{Timeline}

On the left-hand side of the table is given a tentative timeline of the experiments. The termination date of instruments marked with a fading gradient corresponds to their current funding expiry date, but many of them have a good probability of being prolonged beyond. Upgrades are foreseen for several of these facilities or their follow-up programs and we have quoted the improved numbers when available. 

\section{The future of multi-messenger astronomy: challenges and opportunities}

The detection and identification of a neutrino transient source is expected to be the next giant leap for the field of MM astronomy. This could be achieved by the 2030s, thanks to theoretical and experimental progress. Phenomenological models already enable us to assess the detectability of sources, to optimize experimental design and direct the follow-up efforts of steerable telescopes. Improvements on this aspect will come from the high-cadence observations, which will help model the interaction backgrounds responsible for neutrino production. 
More detailed constraints on source content and structure require finer microscopic modeling with numerical methods, that are being developed, with increasingly complete and self-consistent codes. 
The main challenges are to account for the wide energy and spatial scales at play, to accurately describe the time variabilities, and to pin down the relevant source properties.

The observational prospects are promising: several HE neutrino observatories are already in construction and should be in full operation by the beginning of the 2030s, drastically improving the sensitivity to transients~\cite{Schumacher:2021hhm}, and with sub-degree pointing resolution for up-going muon neutrinos. At UHE, a first generation of experiments with the sensitivity to detect the first UHE neutrinos is planned to be fully deployed around 2025-2030. One may note that current experiments such as ARA will remain competitive with this generation. It is not expected that these instruments reach sub-degree angular resolution, except for BEACON ($0.3^\circ-1^\circ$ resolution) and for the first sub-array for the GRAND experiment (GRAND10k), corresponding to $1/20$th of the full array, which will have few arcminute pointing accuracy. The next generation, to be completed in the 2030s, will be better armed for neutrino astronomy, thanks to an order-of-magnitude better sensitivities, to detect (possibly stacked) short and longer transients, and with a handful of experiments achieving sub-degree angular resolution. For several of these instruments, technical challenges, in terms of air-shower reconstruction and radio detection techniques, will have to be overcome for successful operation. This will require experimental prototyping, cross-calibration between various detection techniques, and common efforts in the community to understand the signals emitted by air-showers arriving with non-standard geometries.
On the EM side, several wide-FoV instruments are planned for 2025 and beyond, with great potential for serendipitous discoveries and excellent neutrino follow-up capabilities, owing to shorter response times and fast slew or survey speeds. 

In the future, with the improvement of EM and neutrino instruments, as well as the development of MM networks, coincident detections and false positive associations will become extremely common. To give a sense of the scale of the problem in its current incarnation the Gold alert stream of IceCube transmits neutrinos with a $\sim 50\%$ probability of being astrophysical. In addition, within the angular uncertainty region of each individual muon neutrino it is common to identify $\sim 10$ optical transients with the ZTF and several Fermi-LAT identified sources~e.g.\cite{2019ATel13160....1S}. The problem of random coincidences will increase in proportion to the sensitivity increase of future  survey and follow up telescopes.
In order to distinguish the most plausible associations, theoretical vetoes --simple analytical criteria that can reject a possible association, e.g.~\cite{Rachen98, Guepin17, Kimura:2017kan, Fiorillo:2021hty} -- could be implemented within alert programs. From an observational point of view, narrowing down the search area in the sky will require a precise reconstruction of the neutrino arrival direction. As mentioned several times, sub-degree pointing will be naturally achieved for a few UHE detectors in the 2030s. 
For cascade events in HE neutrino telescopes, machine learning techniques are being explored to improve the angular resolution~\cite{2021JInst..16P7041A, 2019EPJWC.20705004D}. 

Machine learning and deep learning will take an important part in the search for transient neutrino events, and more generally in transient multi-messenger astronomy \cite{Allen19}. Techniques such as convolutional neural networks and graph neural networks have recently attracted attention, due to their ability to extract and treat spatial features in data \cite{Szadkowski:2014zna, Szadkowski:2017qoy, Huennefeld:2017pdh, IceCube:2018gms, DeSio:2019lcr, Garcia-Mendez:2021vts, Reck:2021zqw, Abbasi:2021ryj, Minh:2021opc}. These methods can be used for different purposes. For instance, they can be used to improve source identification using multi-wavelength or MM data \cite{Fraga21, Vicedomini21, Lopez20}. Moreover, these techniques can serve to perform joint analysis of MM signals, for instance by combining time series and images in neural networks, to reconstruct the source properties \cite{Cuoco21}. Machine learning is also used to increase the sensitivity of detectors and improve real time searches \cite{George18}, for instance by detecting GW signals earlier before the compact object merger and thus improve alerts to perform multi-messenger observations \cite{Baltus21, Yu21}. To summarize, these techniques show promising results, with improvements in the treatment of noisy data sets, signal classification, and thus are promising for the reconstruction of neutrino properties, the identification of neutrino point sources and coincident detections.

Synergies and communication will be key to discoveries in the field of transient MM astronomy: between theory and observations, and between the different instruments. 

\section*{Acknowledgements}

Many thanks to our numerous colleagues for providing detailed input on the cited experiments, sometimes performing new calculations and discussing with us about the relevant numbers to quote: Markus Ackermann, Jaime Alvarez Mu\~niz, Carlos Arguelles, Jean Ballet, Didier Barret, Michel Boer, Sara Buson, Mauricio Bustamante, Ken Chambers, Alexis Coleiro, Alan Coleman, Amy Connolly, Frédéric Daigne, Cosmin Deaconu, Mathieu de Naurois, Krijn de Vries, Damien Dornic, Christian Haack, Clancy James, Albrecht Karle, Mansi Kasliwal, Claudio Kopper, Marianne Lemoine, Vladimir Lipunov, Olivier Martineau-Huynh, Miguel Mostafa, Nepomuk Otte, Nahee Park, Luigi Piro, Simon Prunet, Elisa Resconi, Andres Romero-Wolf, Marcos Santander, Makoto Sasaki, Lisa Schumacher, Fabian Sch\"ussler, Ben Shappee, Robert Stein, Olga Suvorova, John Tonry, Michael Unger, Nick van Eindhoven, Aaron Vincent, Stephanie Wissel, Josh Wood, Yi Zhang. We thank Alexis Coleiro, Shigeo Kimura and Fabian Sch\"ussler for insightful feedback that helped improve the manuscript. 

C.G. is supported by the Neil Gehrels Prize Postdoctoral Fellowship. KK is supported by the APACHE grant (ANR-16-CE31-0001) of the French Agence Nationale de la Recherche.

\section*{Author contributions}
The authors contributed equally to all aspects of the article. 

\section*{Competing interests}
The authors declare no competing interests. 


\begin{landscape}
\begin{table}[pht]
\includegraphics[width=\linewidth]{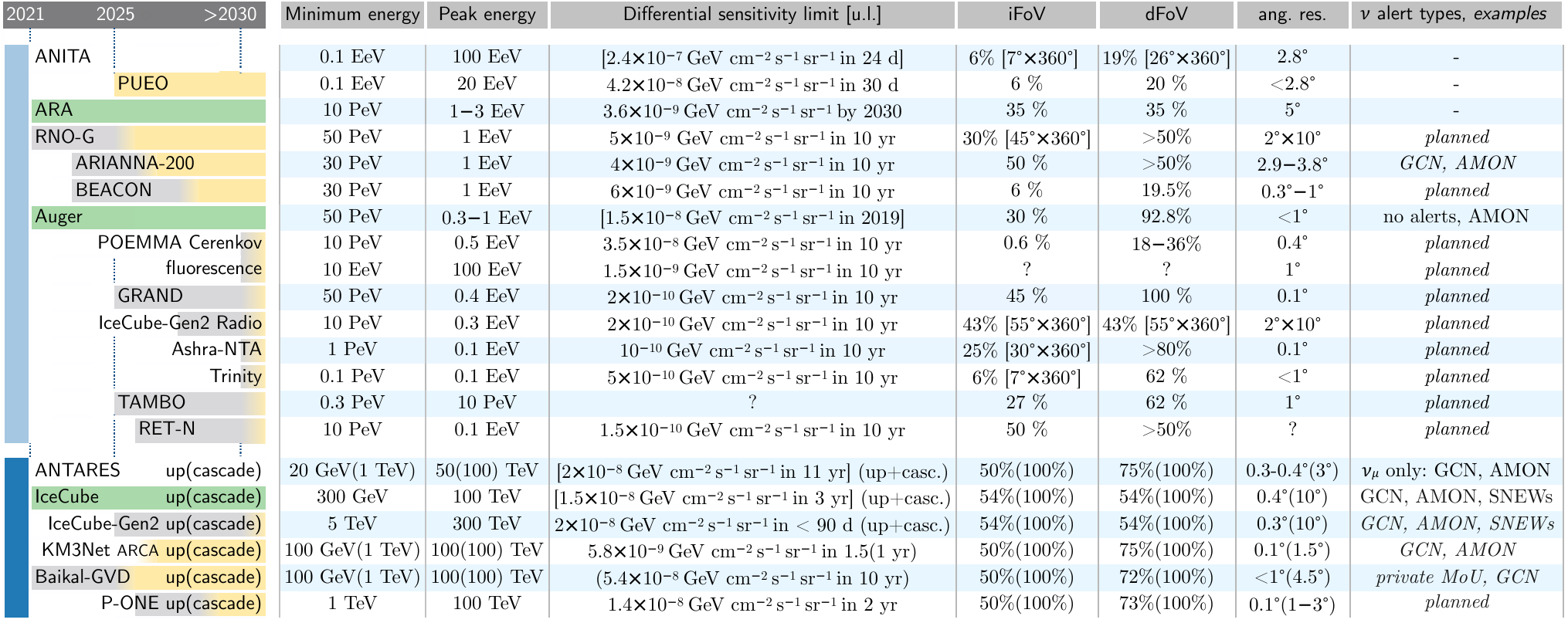}
\caption{Indicative experimental characteristics of current and future neutrino detectors targeting energies above (light blue band) and below (dark blue band) $10\,$PeV. The left-hand side of the table indicates the timeline of each instrument (green: current, yellow: up-coming, gray: under construction). The following columns from left to right reference the minimum neutrino energy, the peak energy where the differential sensitivity is best, the differential sensitivity to diffuse neutrino flux [or measured flux or measured upper limits in brackets], the instantaneous (iFoV) and daily averaged (dFoV) fields of view in sky percentage and in square degrees in brakets, and the angular resolution. The final column provides information on alert programs set up or planned to be set up (in italics) by the instrument. For instruments targeting $<10\,$PeV energies, the numbers in parenthesis are for  `cascade events' (see text for definition), and the others for muon tracks, unless otherwise indicated. Question marks indicate the yet unknown values for up-coming experiments. References are given in the text.}
\label{table:nu_instr_char}
\end{table}
\end{landscape}

\begin{landscape}

\begin{table}[ht]
\includegraphics[width=\linewidth]{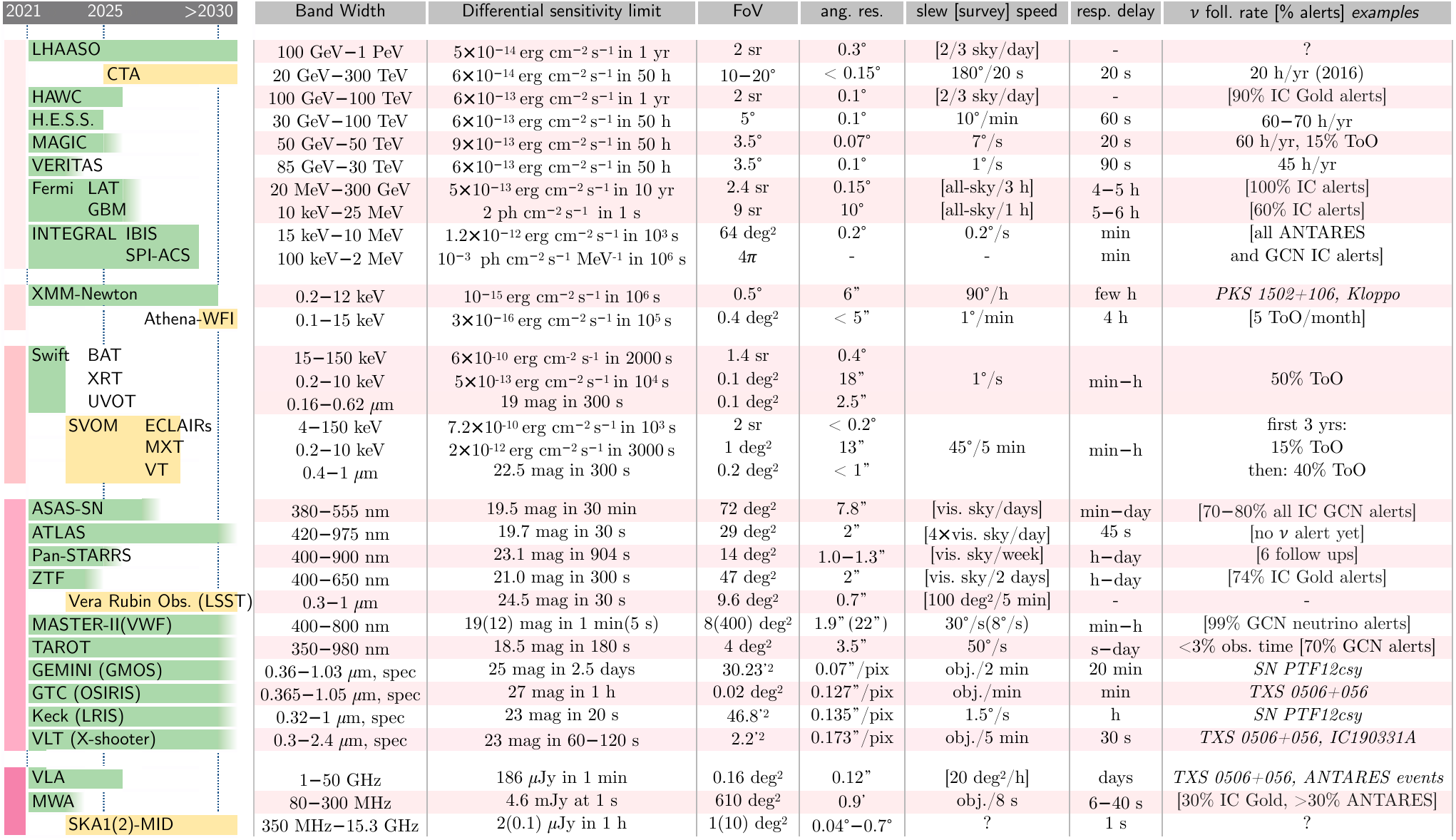}
\caption{Indicative experimental characteristics of a non-exhaustive list of actual or potential neutrino follow-up EM instruments. The left-hand side of the table indicates the timeline of each instrument (green for current and yellow for up-coming). Unclear termination dates are indicated with a fading gradient. The following columns from left to right reference the band width (characterized by either energy, wavelength or frequency range, depending on conventions), the differential sensitivity limit (definition depends on the type of instrument, see \ref{section:EM_diff_sens}), the field of view (FoV), the angular resolution, the slew speed and survey speed in brakets, the response delay to a neutrino or ToO alert. The final column provides elements of the neutrino or ToO follow up program of each facility, with a neutrino alert follow up rate (``$\nu$ foll. rate", in hour/year) when available, percentage or number of neutrinos followed in brakets, and specific followed source or event names in italics. Question marks indicate the yet unknown values for up-coming experiments. References are given in the text.}
\label{table:instr_char}
\end{table}

\end{landscape}

\newpage

\bibliography{NatRev}

\end{document}